\documentclass[twoside]{article}
\usepackage{fleqn,espcrc2}
\usepackage{epsfig}

\def\beq{\begin{equation}}
\def\eeq{\end{equation}}
\def\bea{\begin{eqnarray}}
\def\eea{\end{eqnarray}}
\def\bq{\begin{quote}}
\def\eq{\end{quote}}
\def\gappeq{\mathrel{\rlap {\raise.5ex\hbox{$>$}}
{\lower.5ex\hbox{$\sim$}}}}

\def\lappeq{\mathrel{\rlap{\raise.5ex\hbox{$<$}}
{\lower.5ex\hbox{$\sim$}}}}

\def\Toprel#1\over#2{\mathrel{\mathop{#2}\limits^{#1}}}

\usepackage[figuresright]{rotating}

\newcommand{\AmS}{{\protect\the\textfont2
  A\kern-.1667em\lower.5ex\hbox{M}\kern-.125emS}}

\title{SUMMARY OF NEUTRINO 2000}

\author{John Ellis\address{Theoretical Physics Division, 
        CERN, \\ 
        CH-1211 Geneva 23, Switzerland}%
	}
       
\begin{document}

\begin{abstract}
\begin{center}
CERN-TH/2000-265~~~~~~hep-ph/0008334
\end{center}
 
Aspects of neutrino physics beyond the Standard Model are emphasized,
including the emerging default options for atmospheric and solar neutrino
oscillations, namely $\nu_\mu \rightarrow \nu_\tau$ and $\nu_e \rightarrow
\nu_{\mu, \tau}$ respectively, and the need to check them, the prospects
opened up
by the successful starts of SNO and K2K, the opportunities for future
long-baseline neutrino experiments and high-energy astrophysical
neutrinos. Finally, comments are made on the road map for realizing the
exciting physics potential of neutrino factories.

\vspace{1pc}
\end{abstract}

\maketitle

\section{Models for Neutrino Masses and Oscillations}

It was in 1989 that LEP measured the number of light neutrino species to
be three, and a milestone of this meeting has been the first detection
of $\nu_\tau$ interactions by the DONUT experiment~\cite{DONUT}.
In this talk, I concentrate on aspects of physics beyond the Standard
Model being revealed by neutrinos, principally associated with
neutrino
masses and oscillations~\cite{NUTeV}. There is no good reason why the
neutrinos should
be massless~\cite{Witten}, since there is no corresponding exact gauge
symmetry coupling
to lepton number $L$ that could forbid their masses, analogously to the
way the $U(1)$ of electromagnetism forbids a photon mass.  A neutrino mass
term $m_\nu \nu \cdot \nu$ involves a $\Delta L = 2$ transition and is
generic in Grand Unified Theories (GUTs)~\cite{Mohapatra}.  However, it
is already possible
in the Standard Model, even without invoking `right-handed' singlet
neutrinos $\nu_R$, if one admits non-renormalizable interactions:
\beq
\frac{1}{M} \nu H \cdot \nu H \Rightarrow m_\nu = \frac{\langle 0|H|0
\rangle^2}{M}~,
\label{one}
\eeq
where $M$ is some heavy mass scale $m_W \ll M \lappeq m_P$, and $H$ is a
Standard-Model Higgs field. 

An interaction of the form (\ref{one}) arises naturally from
renormalizable interactions within a GUT, where one
expects a seesaw structure
\bea
(\nu_L, \nu_R) 
\left( \begin{array}{ll}
0 & M_D\\
m^T_D & M
\end{array}
\right) \quad \left( \begin{array}{c}  \nu_L\\
\nu_R \end{array} \right)
\label{two}
\eea
for neutrino masses, where $m = {\cal O}(m_q)$ or ${\cal O}(m_\ell)$,
leading to
\beq
m_\nu = m_D \frac{1}{M} m^T_D~,
\label{three}
\eeq
which is naturally small if $M \sim M_{\rm GUT}$. For example, with $m =
{\cal O}(10)$~GeV and $M = {\cal O}(10^{13})$~GeV one finds $M_\nu =
{\cal O}(10^{-2})$~eV. It is important to 
note, though, that the gauge group of the GUT does not play an essential
r\^ole.

Any such mechanism may be expected to lead to mixing in generation space:
\beq
V_{\rm MNS} = V_\ell V^t_\nu~,
\label{four}
\eeq
where $V_\ell$ diagonalizes the charged-lepton mass matrix, and $V_\nu$
the
neutrino mass matrix (\ref{three}).  {\it A priori}, $V_\mu$ would
originate either in
the Dirac mass $m$ or in the heavy singlet mass matrix $M$.  Since the
mechanism generating neutrino
masses and mixing is different from that for quarks, and more
complicated, perhaps one should not be surprised if the patterns of masses and
mixing are different, too.

An extreme variation on the seesaw mechanism is the way neutrino
masses may arise in models with extra
dimensions~\cite{Dienes}.  For each flavour, one could have an infinite
set of Kaluza-Klein
excitations contributing via mixing to a generalized seesaw:
\bea
\left( \begin{array}{lllll}
0 &\vdots & m_1 & m_2 &\cdots \\
\multispan5 \dotfill \\
m_1 &\vdots & M_1 & 0 &\\
m_2 & \vdots & 0 & M_2 & \\
\vdots & \vdots & &  &\ddots \end{array}
\right)
\label{five}
\eea
In this model, the diagonal entries $M_i$ need not be very large, because
the off-diagonal
entries $m_i$ may be suppressed by the large size(s) of (one of) the extra
dimension(s).  A curiosity of this picture is that the neutrino oscillation
pattern is no longer simply sinusoidal, emphasizing how essential it will be to
measure the oscillation pattern experimentally.

A possibility intermediate between (\ref{five}) and the simple GUT model
(\ref{two}) is when there are $3 < N < \infty$ singlet states~\cite{ELLN} 
mixing in the
seesaw mass matrix:
\bea
\left( \begin{array}{lr}
0_{3 \times 3} & m_{N \times 3} \\
m^T_{3 \times N} & M_{N \times N} \end{array} \right)
\label{six}
\eea
This offers more possibilities for large mixing, e.g., in the case that 
$M_{N \times N}$ is
`dense' with many similar entries, just three entries
in
$m_{N \times 3}$ of comparable magnitudes would suffice to give three
similar light neutrino masses with large
mixing.

\section{Upper Limits on Neutrino Masses}
As we heard at this meeting, the end-point of Tritium $\beta$ decay now provides
the upper limit~\cite{Troitsk,Mainz}
\beq
m_{\nu_e} < 2.2~{\rm eV}~,
\label{seven}
\eeq
there is no longer a problem with $m^2 < 0$, and there are prospects to reach a
sensitivity to $m_{\nu_e} < 0.5$~eV.  The puzzle of the Troitsk seasonal anomaly
remains~\cite{Troitsk}, but we learnt here~\cite{Mainz} that it is not
supported by the
Mainz
experiment~\cite{Gatti}. From $\pi
\rightarrow \mu \nu$ decay we have~\cite{PDG}
\beq
m_{\nu_\mu} < 190~{\rm keV}~,
\label{eight}
\eeq
and there are ideas how this might be improved by a factor $\sim 20$ using
the BNL $(g-2)_\mu$
experiment.  From $\tau \rightarrow (n \pi)_\nu$ decay we now
have~\cite{Roney}
\beq
m_{\mu_e} < 15.5~{\rm MeV}~,
\label{nine}
\eeq
and there are prospects at $B$ factories to improve this by a factor $\sim 5$,
closing definitively the cosmological window for a decaying massive $\nu_\tau$. 
Neutrinoless double-$\beta$ decay experiments~\cite{Ejiri} currently
establishes the upper limit
\beq
\langle m_\nu \rangle_e \equiv \sum_i m_{\nu_i} V^2_{e_i} \lappeq 0.2~{\rm eV}~,
\label{ten}
\eeq
and there are prospects to improve the sensitivity to $\langle m_\nu \rangle_e
\sim 0.01$~eV~\cite{Fiorini}.  As discussed later, the limit (\ref{ten})
is already
causing
problems for models with degenerate massive neutrinos, and this improved
sensitivity could impact many more~\cite{KS}.

\section{Quo Vadis LSND ?}
As we heard here~\cite{LSND}, a re-analysis of the full LSND data set
confirms their
previous results for both decays at rest (32.7 $\pm$ 9.2 events with $R-\gamma >
10$) and in flight.  Their data are neither confirmed nor excluded by
KARMEN~\cite{Eitel},
whose timing anomaly has not reappeared in their latest data set.  It is
desirable to repeat the previous global analysis~\cite{Eitel2} of LSND and
KARMEN
data
using their likelihood functions in the $(\sin^2 \theta, \Delta m^2)$ plane, so
that future experiments know what target region in this plane to aim at.

A definitive test of LSND will be made byMiniBooNE~\cite{Bazarko}, working
at similar $L/E$
but with $E \sim 1$~GeV and starting at the end of 2001.  If needed, this may
be followed up by the full BooNE experiment.  There is also an opportunity to
pursue the LSND effect still further with the ORLaND project at the
SNS~\cite{Avignone}, so I
conclude that there is adequate follow-up of the LSND experiment.

If confirmed, the LSND data would require the presence of a fourth,
sterile light neutrino species $\nu_s$ which should mix with either
atmospheric and/or solar neutrinos.

\section{Atmospheric Neutrinos}
Recent experiments are providing improved understanding of the
cosmic-ray neutrino `beams'.
 AMS~\cite{AMS} and BESS~\cite{BESS} provide improved measurements of the
primary flux, balloon
 experiments constrain shower models~\cite{Lipari}, and the HARP
experiment~\cite{HARP} will provide
 improved data on $\pi$ and $K$ production.  Three-dimensional shower
 simulations are now available~\cite{Lipari}:  first studies indicate that
the conclusions
 drawn from one-dimensional simulations are robust, and that any bias in the
 previous inferred value of $\Delta m^2$ is within the quoted error
range~\cite{Sobel}.
 
\begin{figure}
\epsfig{figure=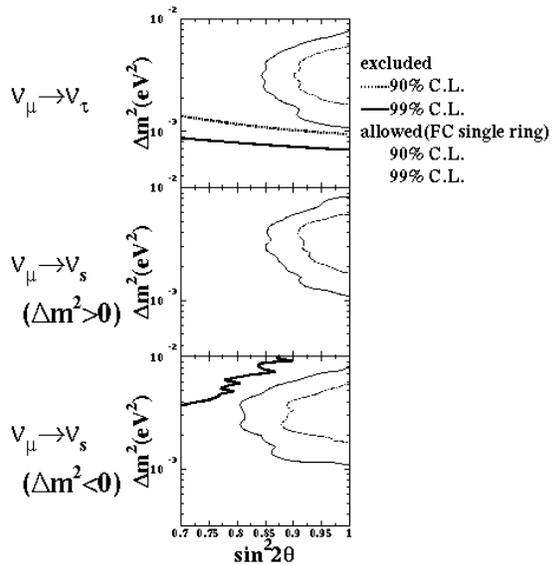,width=7.5cm}
\caption{\it Domains of the $\sin^2 \theta, \Delta m^2$ plane
allowed by the Super-Kamiokande analysis of fully-contained (FC)
single-ring events and excluded by the zenith-angle distributions.
They are compatible for the $\nu_\mu \rightarrow \nu_\tau$ hypothesis
(top panel), but not for the $\nu_\mu \rightarrow \nu_s$ hypothesis
(bottom two panels)~\cite{Sobel}.}
\label{fig:atmo}
\end{figure}

 The impressive statistics on atmospheric neutrinos from
Super-Kamiokande~\cite{Sobel} and
 other experiments~\cite{Soudan,MACRO,Baksan} provide plenty of smoking
guns, in the
forms of zenith-angle
 distributions, up-down asymmetries, etc..  It is clear that something
happens to
 atmospheric $\nu_\mu$ as a function of $LE^n:  n \approx -1$.  Interpreted in
 terms of neutrino oscillations, Chooz~\cite{Chooz} and Palo 
Verde~\cite{Gratta,Mikaelyan} tell
us that
 $\sin^2\theta_{\mu e} \lappeq 0.1$, and Super-Kamiokande favours $\nu_\mu
 \rightarrow \nu_\tau$ over $\nu_\mu \rightarrow \nu_s$ at the 99\% confidence
 level, based on separate analyses of the zenith-angle distributions for
 neutral-current-enriched multi-ring events, partially-contained events with $E
 \lappeq 5$~GeV and upward-going muons, as seen in Fig.~\ref{fig:atmo}.
The rate of $\pi^0$ production may also
 favour $\nu_\mu \rightarrow \nu_\tau$ over $\nu_\mu \rightarrow \nu_s$,
but this needs to be confirmed by data from the K2K near detector, to
reduce
 systematic uncertainties. 

 However, these conclusions should still be taken with a pinch of salt.
The new
 atmospheric-neutrino data should be analyzed for a mixture of $\nu_\mu
 \rightarrow \nu_\tau$ and $\nu_\mu \rightarrow \nu_s$
oscillations~\cite{Betal}, and in
 particular to determine to what extent four-neutrino scenarios motivated by the
 LSND experiment can be accommodated.  The silver bullet would be direct
 evidence for $\tau$ production:  Super-Kamiokande is trying, but may find it
 difficult.  At a more fundamental level, there is as yet no direct
evidence of an
 oscillation pattern~\cite{Lisi}.
 
 These lacunae provide elbow-room for theorists to speculate about more neutrino
 species and/or novel neutrino dynamics.  Neutrino decay and decoherence
 scenarios have both been considered, and appear indistinguishable with present
 data. On the other hand, any significant violation of the principle of
equivalence seems excluded
 by the energy dependence $(LE^n:  n = -1.06 \pm 0.14)$ of the $\nu_\mu$
 deficit~\cite{Lisi}.
 
 These lacunae also provide motivations for future experiments, e.g., to improve
 the sensitivity to $\nu_\mu \rightarrow \nu_e$ oscillations, to look directly
 for $\tau$ appearance, and to measure an oscillation pattern.  They also
 motivate the long-baseline accelerator experiments to be discussed later.  In
 some cases, it may be possible to use the same detector for both accelerator
 and atmospheric `beams':  one exanple is Super-Kamiokande,
MONOLITH~\cite{Geiser} and OPERA
 could be others.
 
 \section{Solar Neutrinos}

\begin{figure}
\epsfig{figure=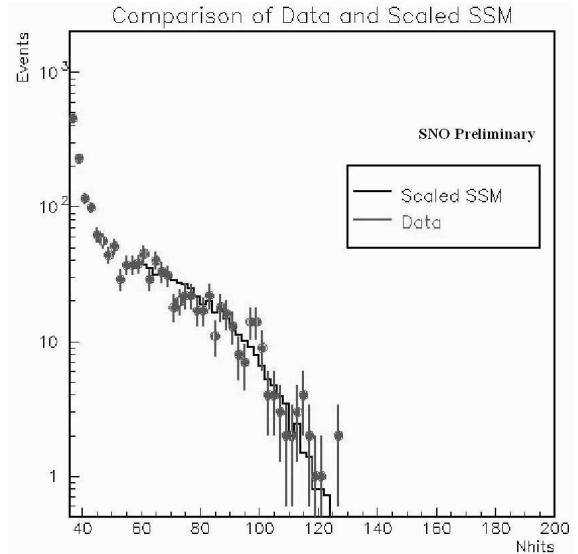,width=7.5cm}
\caption{\it The energy spectrum of CC events observed by SNO~\cite{SNO},
compared with the SSM. The relative normalization is not yet specified.}
\label{fig:SNOE}
\end{figure}

\begin{figure}
\epsfig{figure=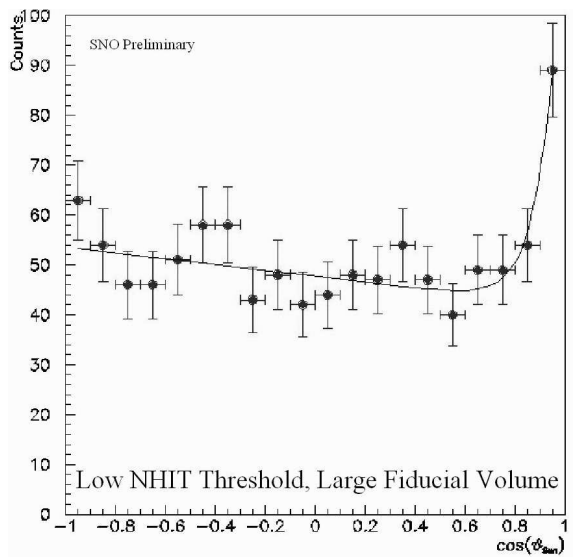,width=7.5cm}
\caption{\it The angular distribution of ES events observed by
SNO~\cite{SNO}.}
\label{fig:SNOangle}
\end{figure}

 We have all been very happy to welcome SNO~\cite{SNO}, the new kid on
this particular
 block~\cite{Homestake}.  SNO has shown us some beautiful events, and they
see a clear signal for
 the charged-current (CC) reactions $\nu_e + d \rightarrow p + p + e^-$,
with a
 clean energy spectrum that seems to be proportional to that of the Standard
 Solar Model (SSM), as seen in Fig.~\ref{fig:SNOE}.  
They also see a clear signal for elastic scattering (ES)
 $\sum_i \nu_i + e \rightarrow \sum_i \nu_i + e$, as seen in
Fig.~\ref{fig:SNOangle}.  

In the present year-long
 Phase I, they plan to measure the ratio of CC and ES rates sufficiently
 accurately to tell us at the 3-$\sigma$ level that $i \ne e$
 high-energy neutrinos are coming from the Sun at the rate expected in
the SSM. Subsequently, for a year in
 Phase II, they will run with salt to measure also the neutral-current (NC)
 cross-section $\sum_i \nu_i + d \rightarrow p + p + \sum_i \nu_i$.  Finally, in
 Phase III they will add $^3He$ detectors to make an independent NC measurement.
In Phases II and III, SNO should be able to discriminate decisively
between different solar-neutrino scenarios~\cite{BKS}.
  
Another newcomer to the solar neutrino field is GNO, which reported here a new
Gallium measurement~\cite{GNO}.  Combined with the previous GALLEX
measurements, their
result is $77.5 \pm 6.2^{+4.3}_{-4.7}$ SNU, far below the SSM.  For comparison,
SAGE now report $75.4^{+7.8}_{-7.4}$ SNU~\cite{Gavrin}. 

Meanwhile, the solar pace continues to be set by
Super-Kamiokande~\cite{Suzuki}. who announced
here data from running for 1117 days, with a lower threshold of 5~MeV and
background reduced
by 60\%.  A re-analysis of all their data yields $0.465 \pm
0.005^{+0.016}_{-0.013}$.  Their energy spectrum is now very consistent with a
constant suppression:  $\chi^2 = 13.7$ for 17 d.o.f.~\footnote{This
impressive agreement is further improved if new measurements of
the Boron $\beta$-decay spectrum~\cite{Boron} are used.}.  

%{\bf Nanie: Espectrum.ps goes here}

\begin{figure}
\epsfig{figure=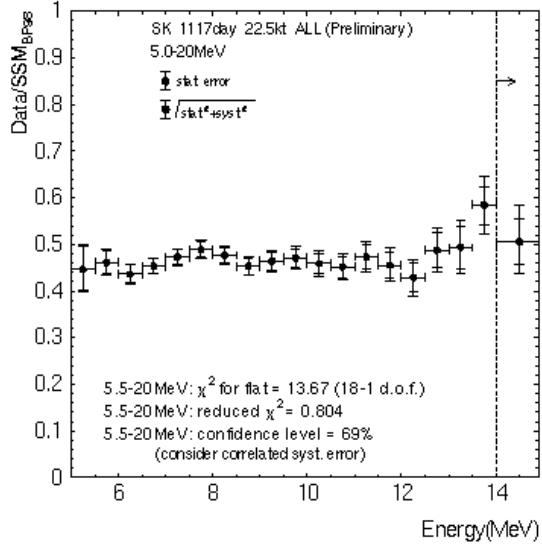,width=7.5cm}
\caption{\it The energy spectrum for solar neutrinos observed by
Super-Kamiokande~\cite{Suzuki}.}
\label{fig:Espectrum}
\end{figure}

They
have established an
upper limit on the hep neutrino flux:  $hep/SSM < 13.2$ for $E > 18$~MeV,
which means
they are essentially irrelevant for global fits.  Super-Kamiokande also report a
reduced day-night asymmetry:  $2(D_N)/(D+N) = -0.034 \pm
0.022^{+0.013}_{-0.012}$, only $1.3\sigma$ away from zero.  

\begin{figure}
\epsfig{figure=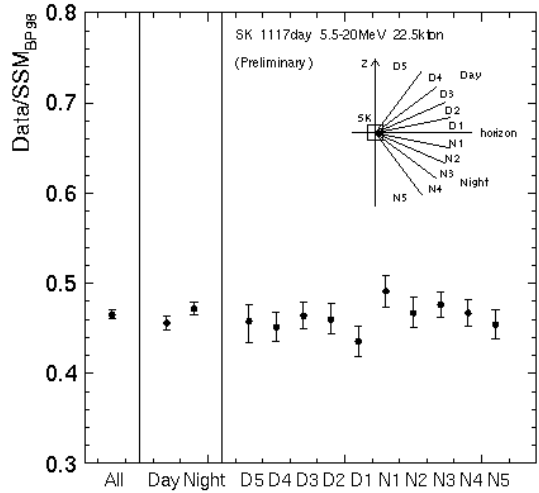,width=7.5cm}
\caption{\it The Super-Kamiokande data on solar neutrinos observed during
different periods of the day and night~\cite{Suzuki}.}
\label{fig:daynight}
\end{figure}

Moreover, the
observed seasonal variation is compatible with the Earth's orbital eccentricity,
and there is no significant correlation with the sunspot
number~\cite{solarmoment}. On the
other hand, as we heard here, the SSM
describes very well the available helioseismological data, indicating that the
true sound speed differs from the SSM by at most 0.002 in the central regions
$R/R_\odot \lappeq 0.1$. Thus it appears difficult to blame the solar
neutrino deficits on a failure of the SSM~\cite{TC}.

The day-night and spectral data from Super-Kamiokande create problems for the
vacuum oscillation (VO) scenario, which they now find to be disfavoured at the
95\% confidence level. They reach the same conclusion for the
small-mixing-angle (SMA) scenario: the day-night data want a larger
value of $\sin^2 \theta$ than is favoured by the spectral data. What
remain are $\nu_e \rightarrow \nu_{\mu,\tau}$
oscillation scenarios with a large mixing angle (LMA) and $\Delta m^2 \gappeq 2
\times 10^{-5}~{\rm eV}^2$, and perhaps the LOW scenario with $\Delta m^2
\sim 10^{-7}~{\rm eV}^2$~\cite{Suzuki}. 

%{\bf Nanie: solarexclude.ps goes here}
\begin{figure}
\epsfig{figure=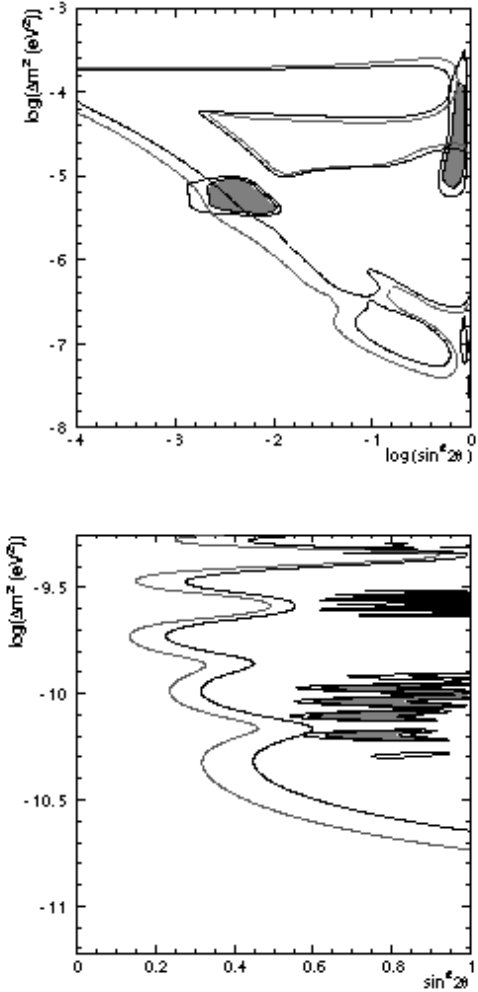,width=7.5cm}
\caption{\it The regions of $\sin^2 \theta, \Delta m^2$ 
for $\nu_e \rightarrow \nu_{\mu, \tau}$ oscillations that are allowed by
analyses of the solar neutrino rates (shaded) and the day-night and
spectrum information. The SMA and VO solutions are excluded at the 95 \%
confidence level, as are $\nu_e \rightarrow \nu_s$
oscillations~\cite{Suzuki}.}
\label{fig:solarexclude}
\end{figure}

However, all these conclusions need to be confirmed by updated global
analyses~\cite{GC,Lisi}. 
These should include unified treatments of the `dark side' with $\theta >
\pi/4$, as well as of matter and vacuum effects.  It is also important to
analyze the data in three- and four-neutrino scenarios.  Even if $\nu_e
\rightarrow \nu_s$ is disfavoured, at what upper limit can such a component be
tolerated~\cite{Betal}?  Conventional interpretations of the LSND
experiment require either
atmospheric $\nu_\mu \rightarrow \nu_s$ oscillations, or solar $\nu_e
\rightarrow \nu_s$ oscillations.  If both these are disfavoured, how can one
interpret LSND?

\section{The Emerging Default Option}
The scenario most favoured~\cite{Kayser} by the current neutrino data
comprises just three
light neutrinos, probably with hierarchical masses (more on this later), and
approximately bimaximal mixing:
\bea
U_{MNS} \simeq \left(
\begin{array}{lcr}
\frac{1}{\sqrt 2} & \frac{-1}{\sqrt 2} & 0 \\ \\
\frac{1}{2} & \frac{1}{2} & \frac{-1}{\sqrt 2} \\ \\
\frac{1}{2} & \frac{1}{2} & \frac{1}{\sqrt 2} \end{array} \right)
\label{eleven}
\eea
The masses of the light neutrinos should be mainly Majorana, they should have
small dipole moments~\cite{moment}, and have lifetimes much longer than
the age of the
Universe.

This list of defaults raises many questions~\cite{Smirnov}.  How may one
rule out light
$\nu_s$?  Can one exclude degenerate neutrinos or an inverse hierarchy of
masses?  How large is $\theta_{13}$?  Are the SMA and VO solutions really
excluded?  How can one discriminate between the LMA and LOW solutions?  Is CP
violation measurable in the neutrino sector?  Can the neutrino mass scale be
fixed by $\beta\beta_{0\nu}$ measurements?  Can the neutrino oscillation
pattern
be seen and decay interpretations of the atmospheric experiments be excluded? 
Finally, how do we actually PROVE that $m_\nu \not= 0$?

\section{Future Solar Neutrino Experiments}
We have seen that the present data disfavour many oscillation scenarios (SMA,
VO, $\nu_e \rightarrow \nu_s$) but there is no `smoking gun' yet:  no spectral
distortion, no day-night effect, and no abnormal seasonal effect.  This provides
an opportunity for SNO (comparing NC, CL and ES rates),
BOREXINO~\cite{Ranucci} and
KamLAND~\cite{Piepke}
(pinning down the $^7Be$ flux).  it also motivates a new generation of
low-energy solar neutrino experiments, to measure the $\nu_e$ energy spectrum
and the NC/CC ratio~\cite{vF}, as discussed here at a pre-conference
workshop~\cite{LowNu}.

\section{Long-Baseline Experiments}
These are needed (a) to convince the remaining sceptics, e.g., by 
using a controllable beam and measuring an
oscillation pattern, (b) to enable precision measurements, e.g., of
$\Delta^2_{m_23}$ and $\sin^2 2\theta_{23}$, and (c) to make new measurements
possible, e.g., of $\tau$ production and $\theta_{13}$.

\begin{figure}
\epsfig{figure=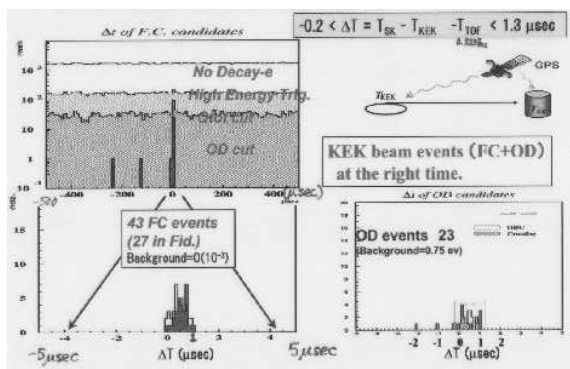,width=7.5cm}
\caption{\it The time structure of events observed by K2K in coincidence
with the KEK beam spill~\cite{K2K}.}
\label{fig:K2Kevents}
\end{figure}

One of the highlights of this meeting was the triumphant start reported by
K2K~\cite{K2K}.
They now have data from $\sim 3 \times 10^{19}$ protons on target: see
Fig.~\ref{fig:K2Kevents}. The beam monitors agree with
Monte Carlo predictions.  Events are seen without significant background in
Super-Kamiokande.  Moreover, K2K is sensitive to the `interesting' range
of $\Delta m^2$ suggested by the atmospheric neutrino data.  

%{\bf Nanie: K2Kevents.ps goes here}

\begin{figure}
\epsfig{figure=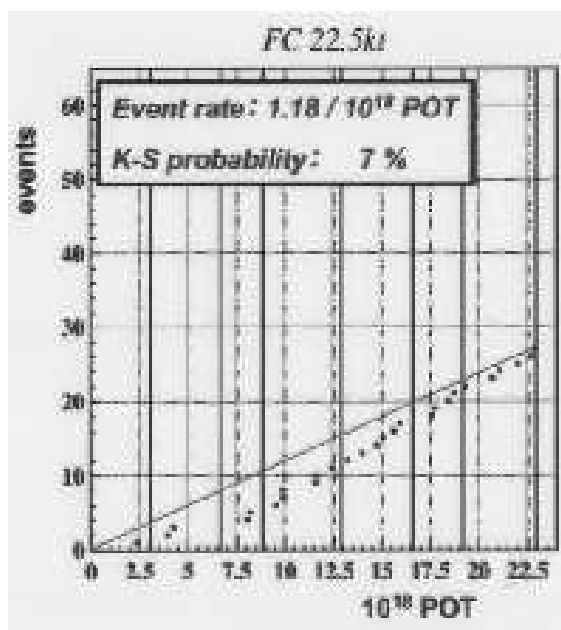,width=7.5cm}
\caption{\it The accumulation of FC events in Super-Kamiokande, as a
function of the number of protons on target~\cite{K2K}.}
\label{fig:K2Ktime}
\end{figure}

Excitingly, K2K reports a promising deficit of events (27 fully-contained 
events in the fiducial volume, compared with $40.3^{+4.7}_{-4.6}$ expected
in the absence
of oscillations), although this is still
only a 2-$\sigma$ effect, and one must beware of the tricks of low statistics. 
The rate of accumulation of events has not been particularly uniform, 
as seen in Fig.~\ref{fig:K2Ktime}, though it is
consistent with statistical fluctuations.  
More data are eagerly awaited, as
well as the analyses of the energy spectrum and of $\pi^0$ production in the
nearby detector. 

Next in line will be KamLAND~\cite{Piepke}, starting in 2001, which should
be able to
test definitively the LMA solution to the solar neutrino deficit, using
neutrinos from nuclear power reactors.  This experiment will also make other
contributions to studies of solar neutrinos, e.g., by measuring the day-night
and seasonal effects for $^7Be$ neutrinos.

MINOS construction is underway~\cite{Woj}.  In addition to confirming
$\nu_\mu$
disappearance, it
should be able to see an oscillation pattern, which will convince
sceptics and enable precision
measurements of $\Delta m^2_{23}$ and $\sin^22\theta_{23}$. MINOS also has
good
sensitivity to $\sin^22\theta_{13}$.  It even has some potential to
discriminate
between $\nu_\mu \rightarrow \nu_\tau$ and $\nu_\mu \rightarrow \nu_s$.
We all
hope there will be no unnecessary delays due to financial problems, etc..

The CERN-Gran Sasso neutrino beam was approved at the end of
1999~\cite{Rubbia}. Its primary
objective is $\tau$ detection, as the `smoking gun' for $\nu_\mu \rightarrow
\nu_\tau$ oscillations.  Two experiments are being proposed for this beam: 
OPERA, which will use an emulsion technique profiting from the experiences
of
CHORUS~\cite{CHORUS} and DONUT~\cite{DONUT}, and ICANOE, which proposes to
use a kinematic reminiscent of
NOMAD~\cite{NOMAD}.  In addition, to $\tau$ observation, experiments in
this beam could have
sensitivity to $\sin^22\theta_{13}$, and the proposed MONOLITH
atmospheric neutrino experiment could also use
the beam for oscillation studies~\cite{Geiser}.

\section{Cosmological Relic Neutrinos}

The numerical abundance of any light neutrino weighing $\ll$1~MeV is independent
of its mass.  Hence its relic mass density $\rho_\nu \propto m_\nu$, and the
simple requirement that neutrinos not exceed the critical density:  $\Omega_\nu
\equiv \rho_\nu/\rho_{\rm crit} \leq 1$ implies that $m_\nu \lappeq 30$~eV. 
Neutrinos in this range would constitute hot dark matter.  A `neutrino' weighing
a few GeV might also have $\Omega_\nu \lappeq 1$, and there is another window
for dark matter particles weighing $> m_{Z/2}$. Either of these two
latter
cases would correspond to cold dark matter, and we return later to searches for
supersymmetric relics that would inhabit the third window~\cite{axion}.

In point of fact, studies of cosmological structure formation imply that
$\Omega_{\rm hot} \lappeq 0.1$, corresponding to $m_\nu \lappeq
3$~eV~\cite{Tegmark}. The
relic neutrino density can be at most comparable to the baryon density: 
$\Omega_b \simeq 0.05$.  The concordance (better than a factor of two) between
the values of $\Omega_b$ extracted from Big Bang nucleosynthesis and
measurements of the cosmic microwave background (CMB) radiation is truly
impressive~\cite{Bond}.  Data on structure formation and the CMB also
suggest that
$\Omega_{\rm cold} \sim 0.3$~\cite{Navarro} and $\Omega_{\rm total} \simeq
1$. The missing
`dark energy' may be a true cosmological constant or slowly-varying vacuum
energy~\cite{Turner}:  $\Omega_\Lambda \sim 0.65$, as indicated
concordantly by data on
large-scale structure,  the CMB and high-redshift supernovae~\cite{SN}.

The question arises how likely it may be that the three light neutrinos are
almost degerate with
\beq
m \sim 2~{\rm eV} \gg \sqrt{\Delta m^2_{\rm atmo}} \gg \sqrt{\Delta m^2_{\rm
solar}}~,
\label{twelve}
\eeq
close to the upper limits imposed independently by tritium end-point
measurements (\ref{seven}) and cosmology.  Any such scenario must respect the
strong constraint (\ref{ten}) from $\beta\beta_{0\nu}$
decay~\cite{Baudis}:
\bea
\langle m_\nu \rangle_e &\simeq&
m \times |c^2_2c^2_3e^{i\phi} + s^2_2c^2_3 e^{i\phi\prime} \\ \nonumber 
&&+ s^2_3
e^{2\phi\prime\prime}| \lappeq
0.2~{\rm eV}~,
\label{thirteen}
\eea
where $m$ is the degenerate neutrino mass, $\theta_2$ and $\theta_3$ are mixing
angles and $\phi, \phi\prime, \phi\prime\prime$ are phases.  The Chooz
experiment~\cite{Chooz} suggests that the third term in (\ref{thirteen})
may be neglected, in
which case the first two must cancel:  $\phi\prime \simeq \phi + \pi$ and
\beq
c^2_2 - s^2_2 = \cos 2\theta_2 \lappeq 0.1 \Rightarrow \sin^2 2 \theta_2
\gappeq
0.99~.
\label{fourteen}
\eeq
Thus maximal mixing is necessary, ruling out the SMA and probably even the
LMA solutions.  If VO is the answer, $\Delta m \sim 10^{10}m$, whereas SMA
and LMA would both require $\Delta m \sim 10^{-5}~m$.  There is a question
whether such extreme degeneracy can be reconciled with renormalization
effects~\cite{EL}.  In models with strictly degenerate neutrinos before
renormalization, there may also be problems with maintaining the required
maximal mixing (\ref{eleven},\ref{fourteen})~\cite{EL}~\footnote{For a
way to evade this question, see~\cite{BR}.}. 

Thus there are theoretical question marks against degenerate neutrino
masses and hence substantial hot dark matter.  However, there are no such
concerns with supersymmetric relics $\chi$ and cold dark matter. 
Constraints from LEP impose $m_\chi \gappeq 50$~GeV~\cite{EFGO}, and the
required
cosmological density $\Omega_{\chi} \sim 0.3$ is found in generic domains
of parameter space~\cite{EHNOS,EFGO}. 

\section{Searches for Cold Dark Matter Particles}

There are three major strategies to search for supersymmetric dark
matter~\cite{MACHOs},
namely via annihilation in the galactic halo yielding $\bar p, e^+$ or
$\gamma$'s~\cite{halo}, annihilation in the Sun or Earth yielding
neutrinos that may
be
detected directly or via muons produced by their interactions in rock
surrounding an underground detector~\cite{trap}, and direct detection of
the elastic
scattering of relic particles within the detector~\cite{GW,Akerib}. 

Indirect searches for annihilation neutrinos via the muons they produce
already appear to rule out some supersymmetric models~\cite{Gondolo}, and
more could be
tested with a 1km$^2$ or
1~km$^3$ detector, although some models would still lie
beyond the reach of such an experiment.  Moreover, one should not forget
neutrino mixing and oscillations, which might alter the flux of muon
neutrinos between their production in relic annihilation and their
detectable interactions~\cite{EFM}. 

There is a recent claim to have observed an annual modulation effect in a
direct search experiment~\cite{DAMA}, which could be due to the elastic
scattering of
supersymmetric relics with 50~GeV $\lappeq m_\chi \lappeq 100$~GeV. 
However, this has not yet been confirmed~\cite{CDMS}, and is difficult to
reconcile
with other experiments' upper limits on the elastic cross
section~\cite{Akerib}. 
Theoretically, the elastic cross section required to reproduce the DAMA
data appears difficult to accommodate in supersymmetric models, unless the
sparticle spectrum is rather unconventional~\cite{EFO}. Future experiments
with
greater sensitivity may be needed to see supersymmetric dark matter
directly~\cite{Zacek}.

%{\bf Nanie: EFO.ps goes here}
\begin{figure}
\epsfig{figure=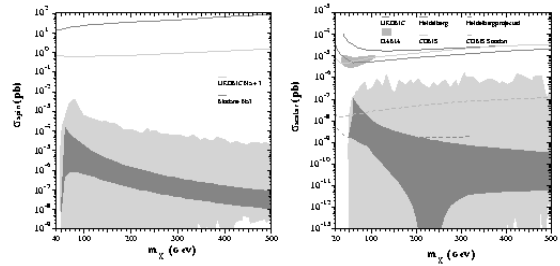,width=7.5cm}
\caption{\it The possible rates of elastic scattering for supersymmetric
relics calculated~\cite{EFO} in the MSSM (light shading) and with
universal scalar masses (dark shading), compared with the region
favoured by DAMA~\cite{DAMA}, the regions excluded by other
experiments~\cite{Akerib} (solid lines), and the possible sensitivities
of future experiments~\cite{Zacek} (dashed lines).}
\label{fig:EFO}
\end{figure}

\section{High-Energy Astrophysical Neutrinos}

Astrophysical neutrinos offer many prospects for probing fundamental
physics as well as possible sources~\cite{Bahcall}.
The first experiments have been taking data for some
time~\cite{AMANDA,Baikal}, and are
beginning to challenge a few models of neutrino emission from AGNs and
GRBs~\cite{Mukherjee,GRBs,Waxman}. New detectors are being
constructed~\cite{NESTOR,ANTARES}, concepts
for the next generation of larger detectors (1~km$^2$ or
1~km$^3$) are being discussed actively, and there are ideas (acoustic
detection, radio detection, observation from space) for increasing further
the detector sensitivity~\cite{Spiering}.  Here I should just like to
mention the
possibility of using high-energy astrophysical
neutrinos~\cite{Ghandi,Weiler}
to probe
fundamental physics~\cite{Scholberg}. 

Based on a couple of approaches to quantum gravity, it has been suggested
that energetic particles might travel at less than the nominal
(low-energy) velocity of light~\cite{AE}: 
\beq
c(E) \simeq c_0 (1 - \frac{E}{m} + \dots)
\label{fifteen}
\eeq
and $\gamma$-ray data from GRBs have been used to set the lower limit $M
\gappeq 10^{15}$~GeV on the possible quantum-gravity scale
$M$~\cite{Mitsou}. Even if
$M \sim 10^{19}$~GeV, it could wash out completely the high-energy $\nu$
pulses that might be emitted by GRBs. 

The modification of relativistic kinematics implicit in (1S) provides one
exotic mechanism for evading the GZK cutoff on ultra-high-energy cosmic
rays~\cite{UHECR}, by forbidding particle production.  There should be
analogous cutoff
for $\gamma$ rays due to the reaction $\gamma + \gamma_{\rm IR}
\rightarrow
e^+e^-$~\cite{UHECR}.  However, TeV $\gamma$ rays from the AGN Mk~501 have
been
reported~\cite{HEGRA}, and the corrected initial $\gamma$ flux would need
to be
extremely high~\cite{PM}.  Moreover, it was shown here~\cite{AMANDA} that
energetic
$\nu$'s from
Mk~501 are not present at a level similar to this corrected spectrum. 
Does this mean that the cutoff has been evaded by a violation
(\ref{fifteen}) of Lorentz invariance? 
We should wait and see, before jumping to such a speculative conclusion!

%{\bf Nanie: Barwick.ps goes here}
 
\section{Neutrino Factories}

For the long-term future of neutrino physics, an attractive concept is a
neutrino factory~\cite{DeRujula,Keil,Schellman}.  Starting with an
intense proton
source -- in the range
of 1 to 20 MW -- with beam energy in the range of a few to 30~GeV, one
hopes to capture and cool $\sim 0.1 \mu /p$, and store them in a `ring'. 
The muons are then allowed to decay, producing $\sim 10^{20}$ to $10^{21}
\nu_\mu$ and $\bar \nu_e /y$ in two or three given directions.  Such a
`ring' may resemble more a bent paper-clip, serving detectors between a
few hundred and a few thousand kilometres away, for different oscillation
studies,
as well perhaps as a nearby detector for other aspects of neutrino
physics. 

%{\bf Nanie: Keil.ps goes here}
\begin{figure}
\epsfig{figure=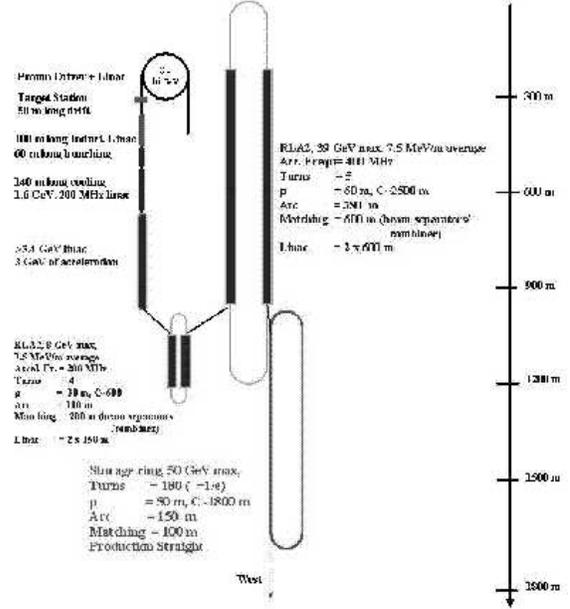,width=7.5cm}
\caption{\it Possible outline design of a neutrino factory~\cite{Keil}.}
\label{fig:Keil}
\end{figure}

Such a facility promises more precise measurements of $\Delta m^2_{23}$
and $\sin^2 2\theta_{23}$, optimal sensitivity to $\theta_{13}$, and the
production and observation of many $\tau$ leptons.  

%{\bf Nanie: theta13.ps goes here}
\begin{figure}
\epsfig{figure=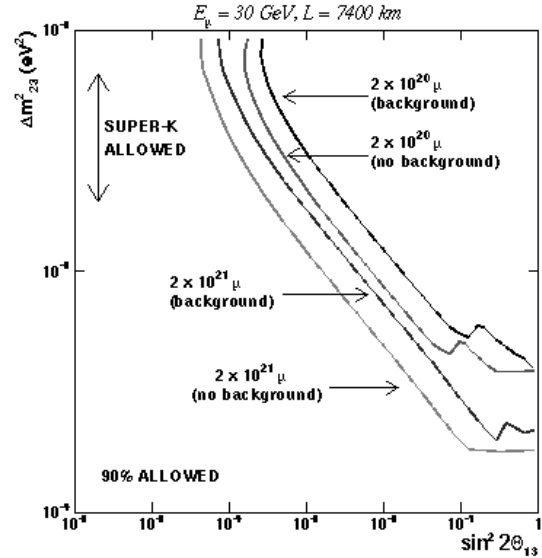,width=7.5cm}
\caption{\it Sensitivity of possible neutrino factory experiments to
to $\theta_{13}$~\cite{DeRujula,Schellman,BCR}.}
\label{fig:theta13}
\end{figure}

The jewel in the crown
of neutrino factory physics~\cite{DeRujula,BCR}, however, could be the
discovery of CP
violation in the neutrino sector~\cite{Volkas}, via the
asymmetry
\bea
A_{\rm CP} & = &
\frac{P(\nu_\mu \rightarrow \nu_e) - P(\bar \nu_\mu \rightarrow \bar\nu_e)}{P(\nu_\mu
\rightarrow \nu_e) + P(\bar\nu_\mu \rightarrow \bar\nu_e)} \\ \nonumber
&&\simeq \frac{4\sin^2\theta_{12} \sin \delta}{\sin\theta_{13}}\sin \left(
\frac{2\Delta m^2_{12}L}{4E} \right)~.
\label{sixteen}
\eea
For this to be observable, large $\Delta m^2_{12}$ and $\theta_{12}$ are
both needed. and both seems to be favoured by the latest Super-Kamiokande
data on solar neutrinos.  There is, however, no good theoretical idea how
large the crucial CP-violating phase $\delta$ might be.  Care must be
taken to disentangle matter effects that could mimic $A_{\rm CP}$, and the
optimal distance for measuring $A_{\rm CP}$ seems to be in the range $\sim
2000$ to 3000~km. Moreover, it has recently been suggested~\cite{Sato} 
that neutrino
CP violation might also be detectable using a conventional low-energy
beam, a possibility that should be reviewed carefully~\cite{Richter}.

%{\bf Nanie: CPnu.ps goes here}
 \begin{figure}
 \epsfig{figure=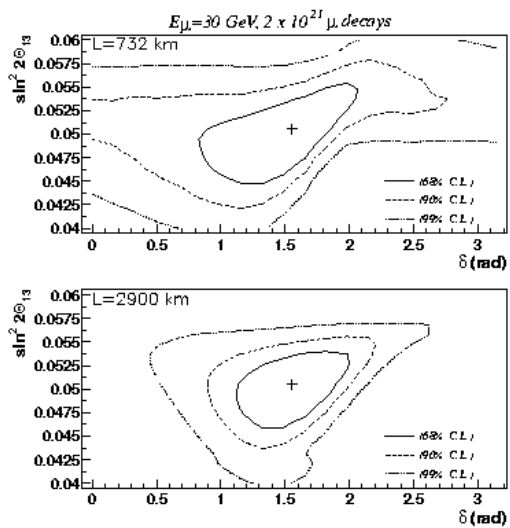,width=7.5cm}
\caption{\it Sensitivity of possible neutrino factory experiments in
the $\delta, \theta_{13}$ plane~\cite{DeRujula,Schellman,BCR}.}
\label{fig:CPnu}
\end{figure}

Moreover, much of the physics
interest of a neutrino factory hinges on the validity of the LMA solution
for solar neutrinos,
which remains to be established. 
In addition to neutrino oscillations, the potential physics cases for
`standard' $\nu$ and $\mu$ scattering physics should be explored, likewise
stopped-muon physics~\cite{raremu}, and perhaps physics with intense kaon
beams~\cite{BCDEGL}, if the
beam energy of the proton source is high enough.

As we were reminded here~\cite{Weiler2}, interest in neutrino physics is
rising monotonically, at least as measured by the relative numbers of
papers with the words `neutrino' and `quark' in their titles. The neutrino
factory is a very exciting project, with the potential to maintain this
rise. It could be a true world machine, involving
experiments in
a different region of the world from the accelerator.  However, it 
remains
a strong intellectual and technical challenge. Building a
neutrino factory will not be a cheap and
small  project. The targeting, muon capture, cooling and 
acceleration, and the engineering of the storage ring itself are all
non-trivial. The proton driver, by itself, is not a large fraction of the
total effort. To realize a neutrino factory, the
neutrino community needs to reach out other communities, 
so as to develop the broadest possible coalition of interested physicists.

%{\bf Nanie: Learned.ps goes here}


\begin{thebibliography}{99}
\bibitem{Baksan}
See, however, S. Mikheyev, for the Baksan Collaboration, talk at this
meeting.
\bibitem{Ranucci}
G. Ranucci, for the BOREXINO Collaboration, talk at this meeting. See also
F.~P.~Calaprice, for the Borexino Collaboration,
Nucl.\ Phys.\ Proc.\ Suppl.\  {87} (2000) 180.
\bibitem{axion}
Coherent axions are another possibility for the cold dark matter:
the prospects for their detection were reviewed here by K.~van~Bibber,
talk at this meeting. See also
L.~J.~Rosenberg and K.~A.~van Bibber,
Phys.\ Rept.\  {325} (2000) 1.
\bibitem{NESTOR}
L. Resvanis, for the NESTOR Collaboration, talk prepared for this meeting.
See also
S.~Bottai, for the NESTOR Collaboration,
Nucl.\ Phys.\ Proc.\ Suppl.\  {85} (2000) 153.
\bibitem{ANTARES}
L. Thompson, for the ANTARES Collaboration, talk at this meeting.
See also
J.~J.~Hernandez, for the ANTARES Collaboration,
Nucl.\ Phys.\ Proc.\ Suppl.\  {81} (2000) 174.
\bibitem{Volkas}
Measuring this might cast some light on leptogenesis scenarios: see
R. Volkas, talk at this meeting.
\end{thebibliography}
\end{document}